\documentclass[twocolumn,a4paper,superscriptaddress,
aps,pra,showpacs,footinbib]{revtex4}
\usepackage{graphicx} 

\newcommand {\mum}{\, \mu \mbox{m}}

\newcommand{\beq}{\begin{equation}}
\newcommand{\eeq}{\end{equation}}
\newcommand{\beqa}{\begin{eqnarray}}
\newcommand{\eeqa}{\end{eqnarray}}
\newcommand{\beqax}{\begin{eqnarray*}}
\newcommand{\eeqax}{\end{eqnarray*}}

\begin{document}
\title{Preparation of atomic velocities by bound-state to resonance conversion}

\author{F. Delgado}
\email{qfbdeacf@lg.ehu.es} \affiliation{Departamento de F\'{\i}sica B\'{a}sica, Universidad de La Laguna, La Laguna, Tenerife, Spain}
\affiliation{Departamento de Qu\'{\i}mica-F\'{\i}sica,
UPV-EHU,\\
Apartado 644, 48080 Bilbao, Spain}
\author {A. Ruschhaupt}
\email{a.ruschhaupt@tu-bs.de}
\affiliation{Institut f\"ur Mathematische Physik, TU Braunschweig,
Mendelssohnstr. 3, 38106 Braunschweig, Germany}
\author {J. G. Muga}
\email{jg.muga@ehu.es}
\affiliation{Departamento de Qu\'{\i}mica-F\'{\i}sica,
UPV-EHU,\\
Apartado 644, 48080 Bilbao, Spain}

\begin{abstract}
A procedure is proposed  
to control the average and width of the velocity distribution
of ultra-cold atoms. The atoms are set initially  
in a bound state of an optical trap  
formed by an inner red detuned laser and an outer blue detuned laser. 
The bound state is later converted into a resonance by a suitable change of
the laser intensities. An optimal time dependence of the switching process,
between the sudden and adiabatic limits, adjusts the final translational
energies to the Lorentzian shape of the resonance state. 
\end{abstract}

\maketitle

\section{Introduction}

With the advent of laser cooling techniques,
the traditional velocity selection or preparation methods \cite{Scoles}
have to be substituted, due to the increasing importance of gravity
and the quantum nature of translational motion, by new methods  
based on   
Doppler sensitive stimulated Raman transitions \cite{chu1, chu2}, 
coherent population trapping
into a dark state \cite{Aspect}, or  
Bragg diffraction by a moving, periodic, optical
potential \cite{SIC99,YYM00,HDK99,CGLV94,KDHW98}.
To complement these methods and overcome some of their limitations 
\cite{KDHW98}, it is worthwhile to explore 
other approaches based on different physical mechanisms.

Fabry-Perot (FP) matter-wave interferometers realized with detuned lasers 
or microwave cavities have been also proposed to provide coherent 
atomic velocity selection or trapping 
\cite{WGTM93,RDM05,LMW98,zhang.1999,martin.2004}.
Moreover, the transmission behavior of a Bose-Einstein
condensate \cite{anglin.2002,bongs.2004} through a double barrier
in a waveguide has been described in \cite{paul.2005}, 
and through an optical lattice in \cite{Car00}.
In a recent paper \cite{RDM05}, we have explored the fundamental limits of a
matter-wave Fabry-Perot optical device made of two blue-detuned laser barriers and a
red-detuned laser well, for selecting both the average and the width of the
atomic velocity distribution. The basic control knob was the well-depth, which   
lets modify the resonance energy.    
It was theoretically and numerically demonstrated that this method may produce arbitrarily small velocities 
but, since it is based on filtering the incident velocity distribution with a resonance 
peak of the transmission probability, the resulting fraction 
of transmitted  atoms 
may be very low and will depend strongly on the incident state.  

The present work describes a modified approach aimed at 
a more efficient preparation and control of the average and width of the final 
velocity distribution. In common with the transmission resonance method, it may also be
implemented with red and blue detuned lasers, 
a red detuned laser forming an inner well and a blue detuned laser for an
outer  barrier. The working principle is now that the atom is initialy
prepared in a bound state. Then, the bound state is converted into
a resonance by a suitable change of
the laser intensities, so the atom will leak out and move
asymptotically with the desired velocity
distribution. This procedure will be descriped in detail in Section
\ref{model}.
A necessary requirement for our method to be of any practical use is
the knowledge of the dependence between laser parameters and the resonances.
This relation will be examined in Sec. \ref{resonances}.
An optimal time dependence of the switching process,
between the sudden and adiabatic limits, will de described in Section
\ref{results} which adjusts the asymptotic kinetic
energies to the Lorentzian shape of the resonance state.
Final conclusions and comments will be provided in Section \ref{conclusions}.

\section{Bound-state to resonance conversion \label{model}}

\begin{figure}
\begin{center}
\includegraphics[height=1.6in,width=2.5in]{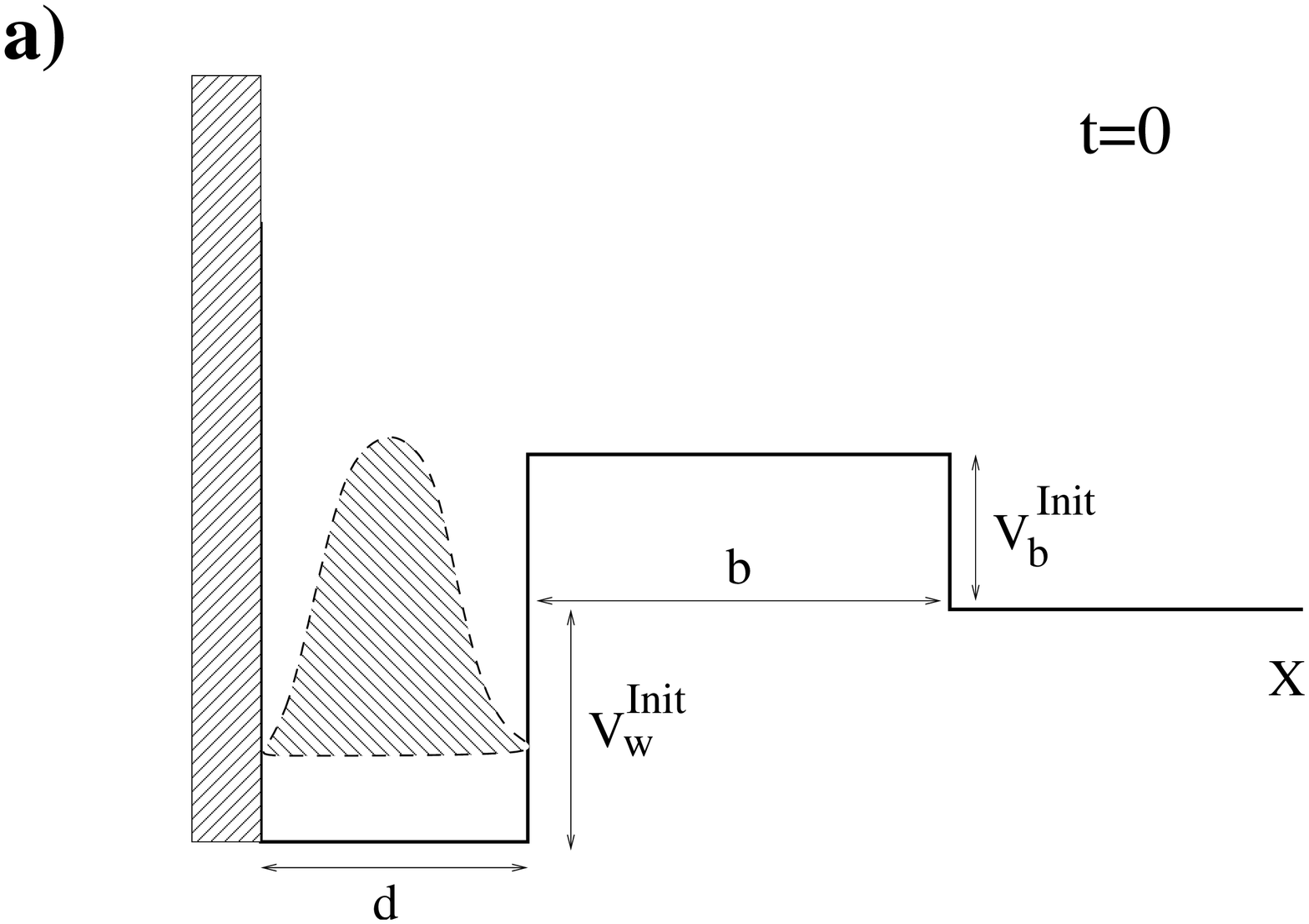}
\includegraphics[height=1.6in,width=2.5in]{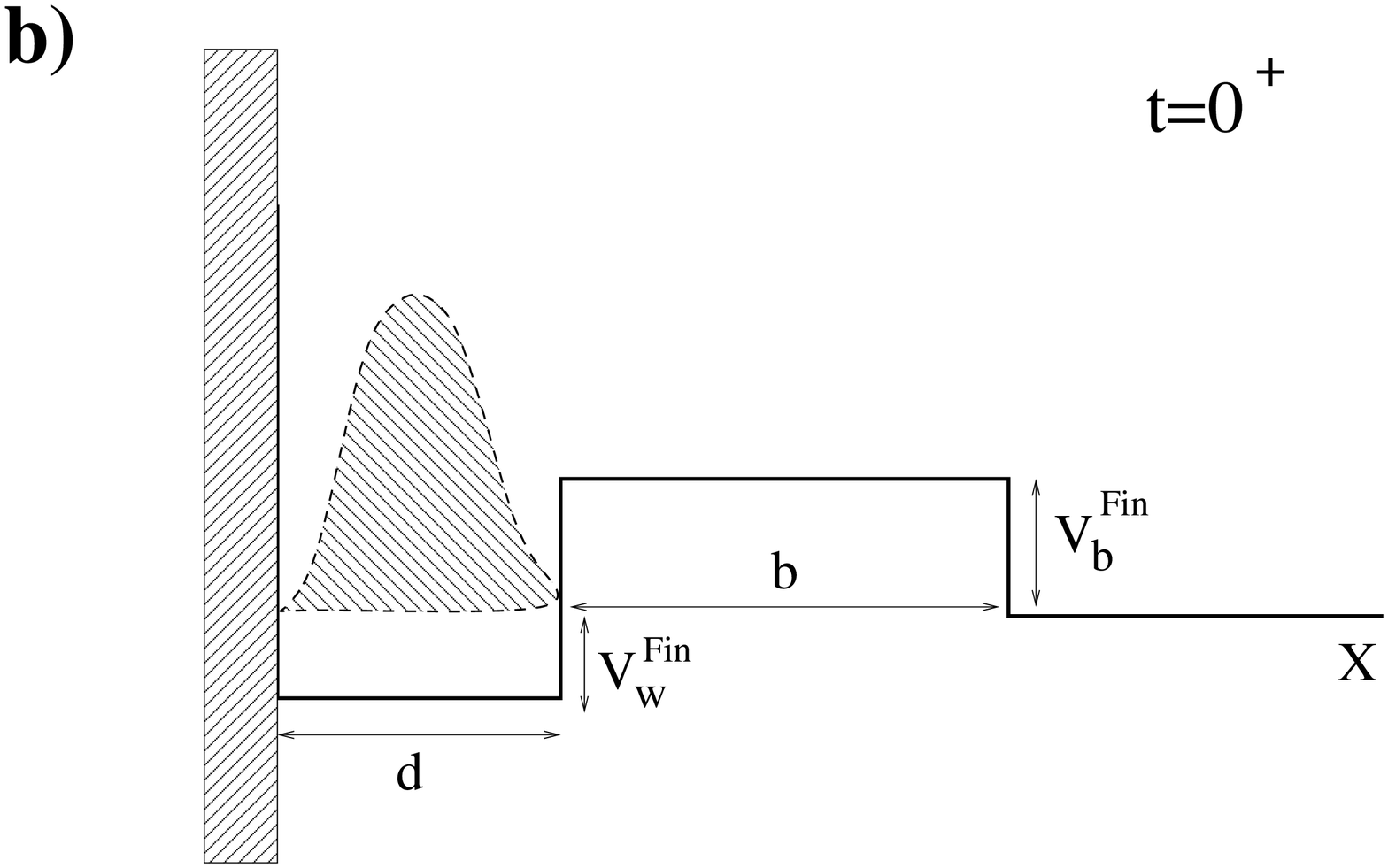}
\includegraphics[height=1.6in,width=2.5in]{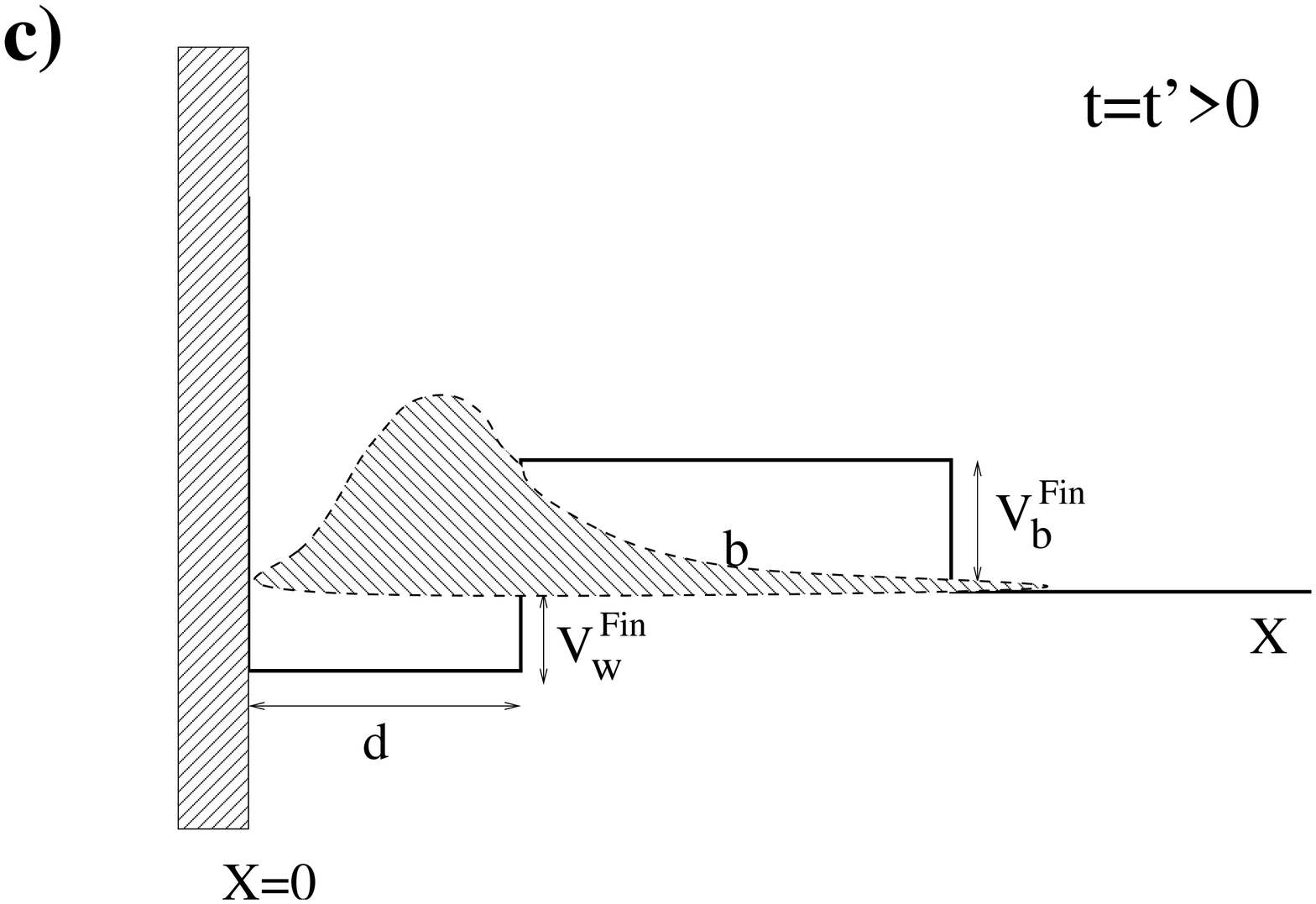}
\end{center}
\caption{\label{fig1} Schematic representation of the velocity preparation method. 
The left barrier represents an ``infinite wall''.   
}
\end{figure}

For simplicity,
we shall assume a one dimensional (1D) model corresponding to the  
effective 1D atomic motion in a narrow waveguide, and a ``square'' shape for each laser intensity, see
Fig. \ref{fig1},
although similar results can be achieved for smoother profiles
\cite{RDM05}. 
The infinite wall at the origin is also a simplifying feature of the 
model, but it is not strictly necessary. In particular, one could 
also use two finite barriers, one at each side of the well \cite{RDM05},
to represent the radial potential profile 
of a cylindrical confinement with free atomic motion or weak confinement in 
the axial direction.
At this stage we also assume a simple single-atom or independent-atoms framework
described by the Schr\"odinger equation and disregard non-linear effects 
that could be incorporated within a mean-field treatment as in \cite{RDM05}.   

The starting point of the velocity preparation  process is a laser configuration 
which holds only one bound state. It is assumed that
the atom can be prepared in this ground state (see Fig. \ref{fig1}(a)).
Several possibilities exist to prepare that initial state: for example, the
original trap could hold more than one bound state; in that case an arbitrary
trapped atomic state overlaps with several of them, but the trap may be
modified to hold one bound state only so that the wave component in the
continuum subspace is eliminated
by its evolution away from the interaction region. More sophisticated and
efficient methods without losing atoms may be based on
ground-state cooling using resolved-sideband transitions \cite{IT3}. Pushing
up the potential well later on, the ground state will eventually become the
only bound state, thus realizing our starting point objective.

Once the initial state of Fig. \ref{fig1}(a) is formed, the potential well is moved
upwards by decreasing the intensity of the red-detuned laser, i.e. $V_w$ is
decreased.
In addition, the intensity of the blue-detuned laser can also be changed.
This is represented in Fig. 1(b), where 
the potential switch has been performed suddenly with respect to other
relevant time scales. 
The consequence is that the bound state becomes, for a final well depth shallower 
than a threshold value, a ``resonant state''. As it is well known, resonances
may be regarded as quasi-bound states associated with poles of the $S$-matrix
in the lower half-momentum plane; they can be linked continuously with bound
states (poles 
on the positive imaginary axis) by varying the potential parameters.  
An important difference though, is that bound states are in Hilbert space and
normalizable, while Gamow 
(resonant) states are not, since they increase exponentially at large
distances from the potential center.  The normalized state achieved by
shifting the well bottom, as in Fig. \ref{fig1}(b),  is thus not a true Gamow
state, but 
it will share approximately some of its properties, in particular its decay
rate, the basic Lorentzian shape in energy space and its coordinate-space form
in the potential
region. We insist that this agreement is necessarily a partial one.

After the switching process, the atom will leak out
(see Fig. \ref{fig1}(c)) having a given (total) energy distribution. Note that
the energy distribution calculated at the end of the switching 
process, i.e., at a time when the atom is still interacting 
with the trap, is equal to the kinetic energy distribution
of the released atoms at asymtotically large time, as it follows from
energy conservation.
Therefore at a sufficient large time, the atom will move with the
desired velocity distribution.

Two limits consisting on sudden or infinitely slow 
well switching may be considered: (a) A sudden well shift produces a state with 
contributions from higher resonances. They will lead to perturbations with respect to the ideal velocity distribution
which will affect the short time decay behavior; 
(b) The opposite limit of very slow switching implies a different problem:  
since the pole motion in the complex momentum plane up to 
the final desired resonance position is slow, a continuum 
of other resonances are excited. They will have a decay time larger than the one desired, thus inducing a deviation with respect to the 
desired exponential decay rate, in this case due to a bias
towards slow components.  
We will see in Section \ref{results} that an adjustment of the switching time
may avoid the perturbations of the fast and the slow processes and produce an
excellent agreement with the Lorentzian shape of the energy distribution.

An implementation of the proposed velocity preparation method will require the
knowledge of the dependence between well/barrier parameters and the resonances
such that the atom leaving the trap will have the desired velocity distribution.
This dependence can be achieved experimentally or theoretically, as it will
be described in the next section.

\section{Configurations and corresponding resonances \label{resonances}}
We consider the model based on a well and a barrier with variable
depth/height represented in Fig. \ref{fig1}. The effective potential is
assumed to take the initial and final forms
\beqax
V^{Init/Fin}(x)= \left\{
\begin{array}{ll}
\,\infty & \; :\,x\le 0\\
-V_w^{Init/Fin} & \;:\, 0< x\le d\\
\,V_b^{Init/Fin} & \;:\,d< x\le d+b\\
\,0 & \;:\,x>d+b 
\end{array} \right.\, .
\eeqax
In the final configuration, see Fig. \ref{fig1}(b), the stationary states of a single ultra-cold
atom moving along the $x$ direction will satisfy 
\begin{eqnarray*}
\left[ -\frac{\hbar^2}{2m}\,\frac{\partial^2}{\partial x^2} 
+  V^{Fin}(x)\right] \psi_k(x)=E_k \psi_k(x),
\end{eqnarray*}
where $E_k=\frac{\hbar^2k^2}{2m}$.
%
For the calculations we have used the mass 
of $^{23}$Na. The scattering states will have the form
\beqax
\psi_k(x)=\frac{1}{\sqrt{2\pi}} \left\{
\begin{array}{ll}
C_1e^{iqx}+C_2 e^{-iqx} &\,:\, 0\le x\le d\\
C_3e^{iq'x}+C_4 e^{-iq'x} &\,:\, d\le x\le d+b\\
e^{-ikx}-S(k)e^{ikx} &\,:\, x\ge d+b 
\end{array} \right.,
\eeqax
where $q=\sqrt{k^2+2mV_w/\hbar^2}$ and $q'= \sqrt{k^2-2mV_b/\hbar^2}$
(for the rest of this section we omit the superscript $^{Fin}$). $q$
and $q'$ have a branch cut in the $p$-plane joining the two branch points
at $\pm i\sqrt{2mV_w/\hbar^2}$ and $\pm \sqrt{2mV_b/\hbar^2}$, respectively.
The different coefficients are obtained from the matching conditions
at $x=0$,
$x=d$, and $x=d+b$. The resonances and bound states can be calculated from
the poles of the $S$-matrix in
the complex $k$-plane. They are  
solutions of the equation
\beqax
\Omega(k)&:=&-(k-q')\big[q+e^{2idq}(q-q')+q'\big]\cr
&+&e^{2ibq'}(k+q')\big[q-q'+e^{2idq}(q+q')\big]=0.
\eeqax           
The corresponding roots in the upper half-imaginary axis are  
the bound
states of the system, while the roots in the fourth and third quadrant are
resonances and antiresonances, respectively.

An alternative way to find the resonances is to look for jumps of the 
phase shift $\delta(k)$,
\beqax
\delta(k)=\frac{1}{2i}\log\left[S(k)\right],
\eeqax
or the peaks of the Wigner delay time $\Delta t$,
\begin{eqnarray}
\Delta t(k)=2\hbar \frac{\partial \delta(E_k)}{\partial E_k}=
\frac{2m}{\hbar k}\frac{\partial \delta(k)}{\partial k}.
\label{dt}
\end{eqnarray}
This may be easier than determining the poles by analytical 
continuation of $S(k)$. 
In the Breit-Wigner regime of isolated and sharp resonances,
the information about the resonances given by 
these peaks may be straightforwardly related to the poles in the 
complex $k$-plane. If $E_{res}=E_R-i\Gamma/2$ is the complex energy obtained
from the phase shift (central position $E_R$ and width
at half height $\Gamma$), then
\begin{eqnarray*}
E_R =\frac{\hbar^2}{2m}\left(k_1^2-k_2^2\right),
\quad
\Gamma =\frac{2\hbar^2}{m}k_1k_2,
\end{eqnarray*}
where $k_{res}=k_1+ik_2$ is the corresponding complex wave-number for which  
$\Omega(k_{res})=0$. 

\begin{figure}
\begin{center}
\includegraphics[angle=-90,width=0.75\linewidth]{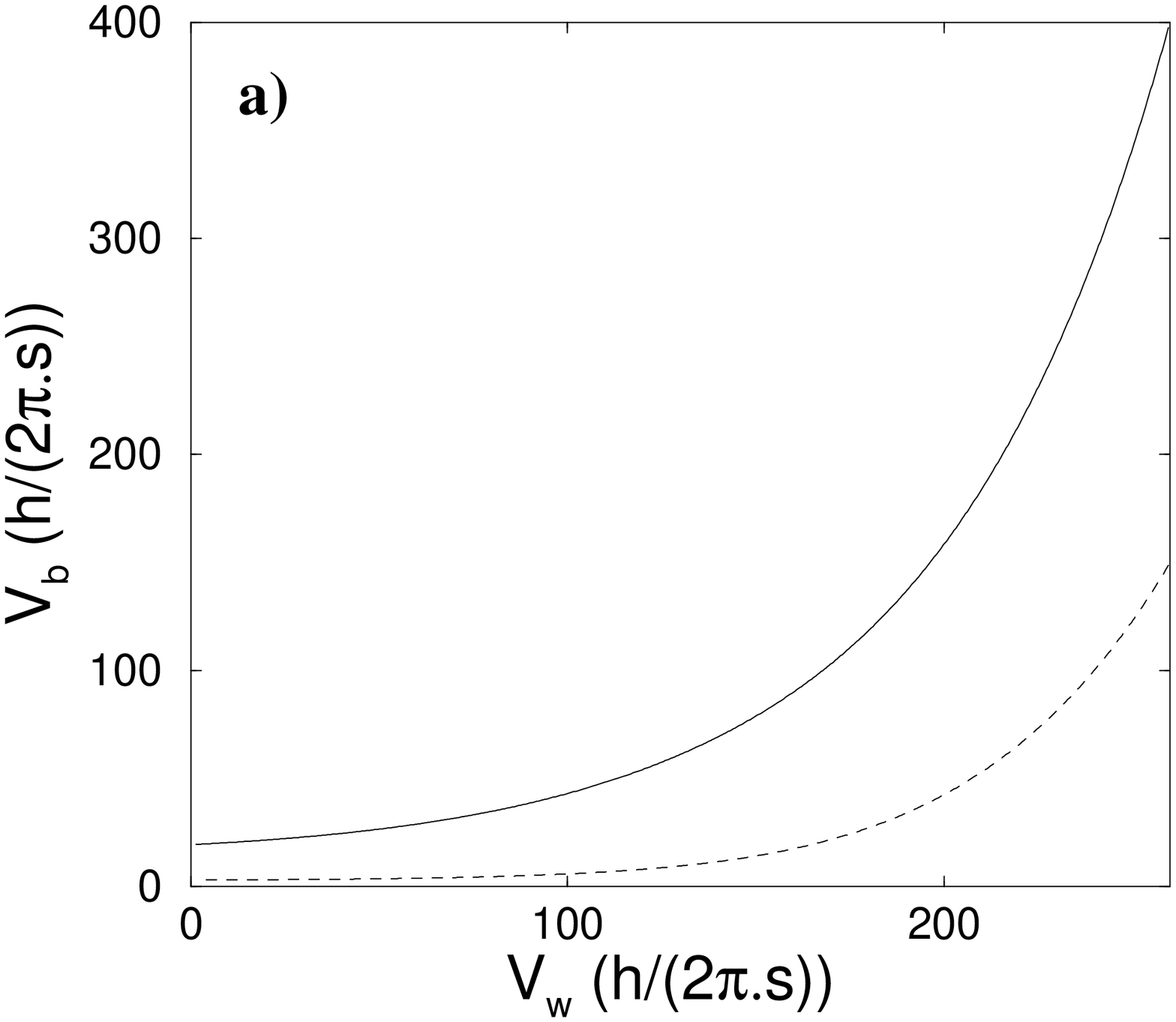}
\includegraphics[angle=-90,width=0.75\linewidth]{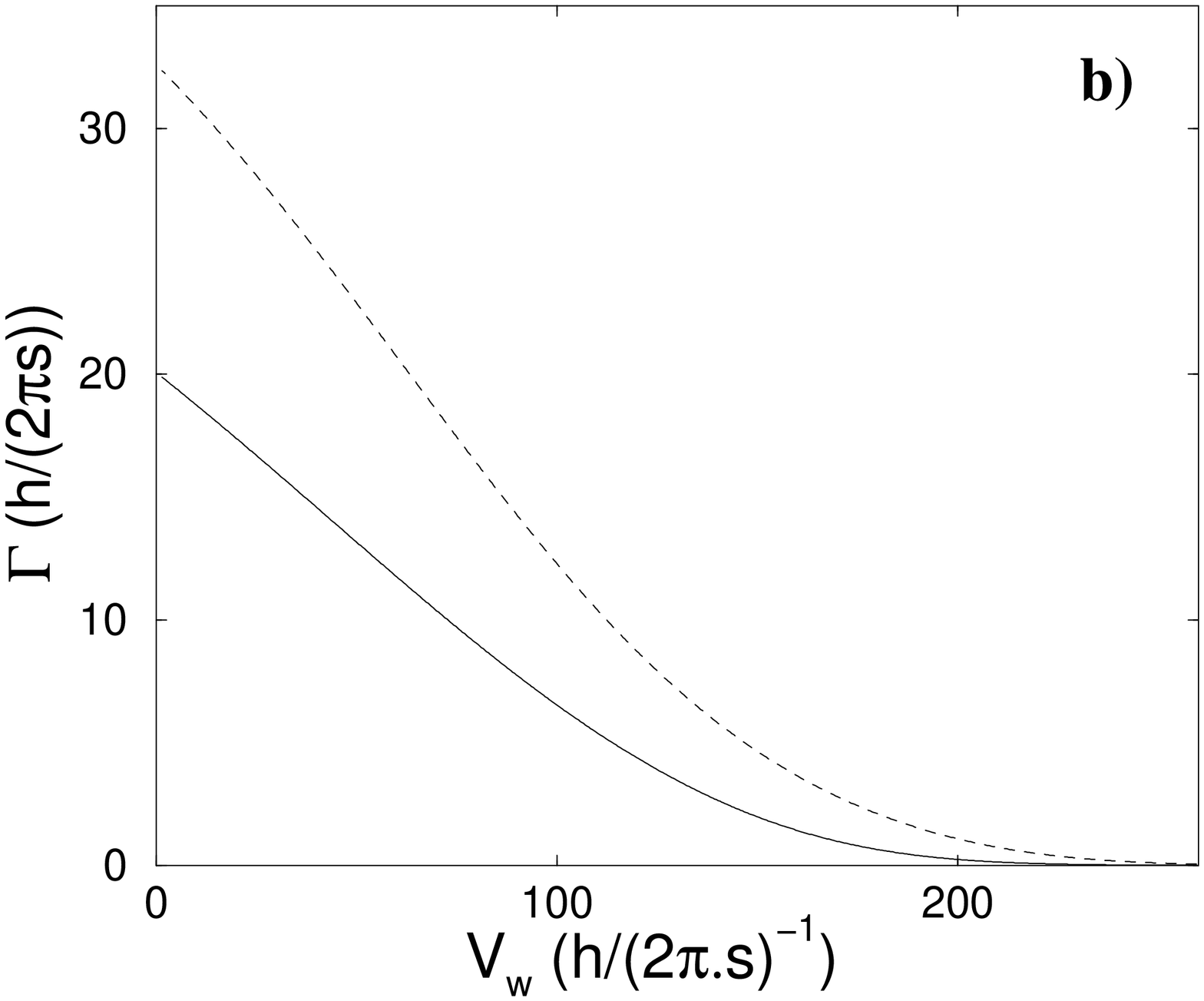}
\end{center}
\caption{\label{fig2} Barrier height $V_b$ versus well depth $V_w$
(upper panel) and, energy width $\Gamma$ versus $V_w$
(lower panel) for constant resonance energies $E_R$:   
$E_R=53.391 \hbar$/s (solid lines) and
$E_R=7.422 \hbar/$s (dashed lines).
$b=10 \mum$, $d=5 \mum$.
}
\end{figure}

Different combinations of well depth $V_w$ and
barrier height $V_b$  lead to the same resonant energy $E_R$.
In Fig. \ref{fig2}(a) we have plotted
the curves for two different resonant energies $E_R$ with
$b = 10 \mum$, and $d=5 \mum$.
Along a curve for a fixed $E_R$, the resonance width will
change. This change is plotted in Fig. \ref{fig2}(b).
Notice that for small
values of the well depth $V_w$, the barrier height $V_b$
that keeps the resonant
energy $E_R$ fixed is almost constant.
In contrast,  $\Gamma$ changes drastically
for small changes of $V_w$ in shallow wells.

\section{Results for different switching times}\label{results}
\begin{figure}
\begin{center}
\includegraphics[angle=-90,width=0.8\linewidth]{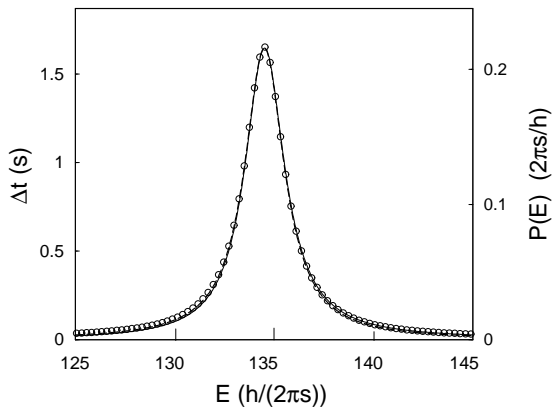}
\end{center}
\caption{\label{fig3} Delay time $\Delta t$ (solid line, left axis), 
versus incident energy of the chosen resonance (see text); the thick dashed line
corresponds to a fitting with a Lorentzian profile.
Energy distribution $P(E)$ (circles, right axis) of the state 
which is initially the ground state for $V_w^{Init}=350\hbar/$s,
 $V_b^{Init}=400\hbar$/s, in  
the final configuration $V_w^{Fin}=100\hbar/$s,
$V_b^{Fin}=200 \hbar/$s, $d=5\mum$, $b=10\mum$.
}
\end{figure}

Let $\varphi_0 (x)$
be the ground state in the initial potential configuration
of Fig. \ref{fig1}(a) with $V_w^{Init}=350\hbar/$s, $V_b^{Init}=400\hbar$/s.
For the destination resonance in Fig. \ref{fig1}(b), we choose the lowest
one corresponding to 
$V_w^{Fin}=100\hbar/$s and $V_b^{Fin}=200 \hbar/$s.
In this case, the resonant complex energy is $E_{res}=(134.509-i1.217)
\hbar/s$. To characterize this resonance we have plotted 
in Fig. \ref{fig3} the delay time $\Delta t$, see Eq. (\ref{dt}),
versus the incident energy. Also shown is the Lorentzian fitting. 

\begin{figure}
\begin{center}
\includegraphics[angle=-90,width=0.75\linewidth]{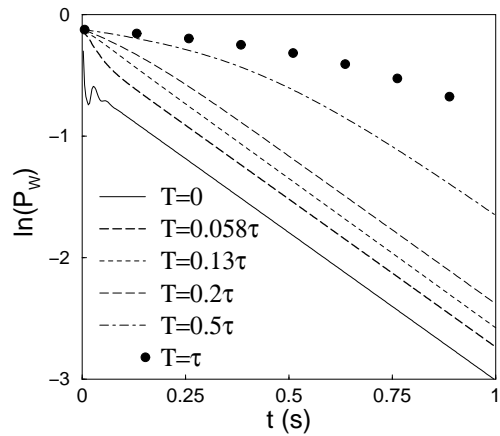}
\end{center}
\caption{\label{fig4} Decay of the probability $P_W (t)$ to find the atom in the well versus time for different values of $T$, see Eq. (\ref{dept});
the initial and final potential configuration are given in Fig. \ref{fig3}.}
\end{figure}

If we move suddenly the bottom of the well making it 
shallower until the initial bound
state overlaps strongly with the desired resonance (Fig. 1(b)), 
the wave-function will evolve in
time, and the atoms will leak out through the barrier (Fig. 1(c)).
We assume that the ensemble of
non-interacting atoms satisfy during this process
the one-dimensional time dependent Schr\"odinger equation,
\beqa
i\hbar\frac{\partial \psi(x,t)}{\partial t}=\left[-\frac{\hbar^2}{2m}
\frac{\partial^2}{\partial x^2}+V^{Fin} (x)\right]\psi(x,t),
\eeqa
where $\psi(x,0) = \varphi_0 (x)$.
As described above, our main objective is to 
achieve a distribution as close as possible to the 
Lorentzian distribution associated with a Gamow resonance
of the final potential configuration (Figs. 1(b) and 1(c)).
Fig. \ref{fig3} shows also the resulting total energy distribution 
$P(E)=\sqrt{\frac{m}{2\hbar^2 E}} |\langle \psi_{k(E)}|\varphi_0\rangle|^2$ of the wavepacket in the new
potential configuration. ($P(E)$ coincides with the kinetic energy distribution when the packet moves away from 
the potential region. \footnote{This is easily derived from the intertwining 
relation of scattering theory, the isometry of Moller operators,
and the assumption that there is no bound state in the final configuration.})   
After such a sudden process, several 
resonances are excited as it may me seen in different ways: note in particular 
that the Lorentzian of Fig. \ref{fig3} is not normalized. This means that 
part of the norm is in higher resonances. A consequence is the fast decay 
of the non-escape probability 
\begin{eqnarray*}
P_W(t)=\int_0^d dx\;|\psi(x,t)|^2
\end{eqnarray*}
at short times
in Fig. \ref{fig4} (solid line). In other words,   
with the sudden switching a significant
fraction of atoms is released 
at early times with ``too much'' energy.
Of course,
if we discard the early, fast atoms,
the decay occurs finally with the desired rate and energy distribution, 
see Fig. \ref{fig4}. However, we may try to produce an ensemble 
without undesired  
high velocity components.   
This can be achieved by a progressive, rather than abrupt, 
switching of the potentials.  

Let us assume that the potential profile changes
in time according to the smooth function
\beqa
V(t,x)=[V^{Fin}(x)-V^{Init}(x)](1-e^{-t/T})+V^{Init}(x). 
\label{dept}
\eeqa
The sudden change corresponds
to $T =0$ and the infinitely slow change to $T=\infty$. 
In Fig. \ref{fig4} we show the decay of the non-escape probability $P_W(t)$.
The lifetime of the first resonance (calculated from the pole of the $S$ matrix)
is $\tau=0.411$ s, in perfect 
agreement with the fitting to the exponential decay that dominates after the 
early transients, independent of $T$. This occurs because 
the final potential configuration is common to all cases so that
the resonance with the
longest life time is the same in all cases.
Nevertheless, the first transient regime varies substantially with $T$, and
for $T\approx\tau$ the initial decay is slowed down considerably.     
The best fit to the purely exponential decay is found for $T\approx 0.13\tau$.

\begin{figure}
\begin{center}
\vspace{0.5cm}
\includegraphics[angle=-90,width=0.8\linewidth]{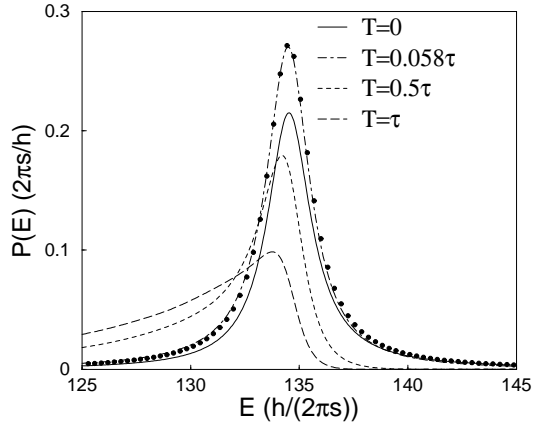}
\end{center}
\caption{\label{fig5} 
Final energy distribution 
for different transition times $T$.
The dots correspond to the 
normalized Lorentzian distribution with the parameters extracted from the
$S$-matrix.
The initial and final configuration are the same than in Fig.
 \ref{fig3}.}
\end{figure}

Our interest is in the
asymptotic and stationary energy distribution at large time, $t_{\infty}\gg T$.
In Fig. \ref{fig5} we have
plotted the final energy distribution
for different transition times $T$. 
For $T=0$ the distribution 
at the main resonance can be approximated by the
normalized Breit-Wigner Lorentzian form corresponding to the 
pole of the selected resonance, however its height is reduced 
because of the 
excitation of higher resonances. 
Increasing $T$, an optimal value is found so that  
the energy distribution fits even in magnitude to the ideal Lorentzian shape.
For the case studied in Fig. \ref{fig5}, the best fit to the Lorentzian of
the selected resonance corresponds to $T \approx 0.058\tau$.
Note that this optimal
value of $T$ is different from the one that provides the 
best fit to the purely exponential decay ($T\approx 0.13\tau$), compare 
Figs. \ref{fig4} and \ref{fig5}.
As $T$ is increased further, the distribution is deformed, the symmetry is lost and
a distortion favoring lower energies is observed.
%

\section{Conclusions and discussion}\label{conclusions}
We have proposed a method to prepare states with well defined average velocity 
and width, based on the conversion between a bound state and a resonance by changing
the trapping potential. 
In atom optics the potentials may be realized with detuned lasers. 
An optimal switching time exists so that the kinetic energy distribution of 
the leaking atoms fits the Lorentzian form of the resonance.  
%

Laser fluctuations may tend to broaden or blur quantum resonances. 
Nevertheless, stabilized lasers provide 
{\em effective constant intensities} in the time scale of $\tau$
and $T$ ($\sim\;0.01-1$ s) \cite{NDGF89,SHT97,RWBTK86} so,
that our analysis would apply to the effective,
time-averaged potentials. Further theoretical and experimental work 
is required to determine the feasibility of using scattering 
resonances in optical traps.    
 
Finally, the method may be applied to electrons 
in semiconductor heterostructures, 
where the well-depth is modified by potential gate voltages
\cite{Austing, Austing2}.

\end{document}